 \documentclass[twocolumn,showpacs,superscriptaddress,aps,prc,10pt,nofootinbib]{revtex4-1} 
 \usepackage[pdftex]{graphicx} 
 \usepackage{cancel} 
 \usepackage[usenames,svgnames,table]{xcolor} 
 \usepackage[english]{babel} 
 \usepackage{ucs} 
 \usepackage[utf8]{inputenc} 
 \usepackage{siunitx} 
 \usepackage[version=3]{mhchem} 
 \usepackage{multirow} 
 \usepackage{textcomp} 
 \usepackage{soul} 
 \usepackage{dcolumn}
 \usepackage{bm}

\newcommand{\ca}{$^{40}$Ca} 
 
\newcommand{\caa}{$^{48}$Ca} 

\newcommand {\amev}{$\,$MeV/nucleon}

\newcommand {\ie}{\textit{i.e.}}

\newcommand {\neos}{nEoS}

\newcommand {\nsuz}{$\langle N/Z \rangle$}
\newcommand {\estara}{$E^{*}/A$}

\usepackage[hyperfootnotes=false,colorlinks=true,allcolors=blue!60!black]{hyperref}

\begin{document} 
%
%
\title{Influence of fast emissions and statistical de-excitation on the isospin transport ratio}

\author{A. Camaiani}  \email{alberto.camaiani@fi.infn.it} 
  \affiliation{Dipartimento di Fisica,  Universit\`a di Firenze, Italy} 
  \affiliation{INFN, Sezione di Firenze, Italy} 
\author{S. Piantelli} 
  \affiliation{INFN, Sezione di Firenze, Italy} 
\author{A. Ono} 
  \affiliation{Department of Physics, Tohoku University, Sendai 980-8578, Japan} 
\author{G. Casini} 
  \affiliation{INFN, Sezione di Firenze, Italy} 
  \author {B. Borderie}
  \affiliation{Universit\'e Paris-Saclay, CNRS/IN2P3, IJCLab, 91405 Orsay, France}
  \author{R. Bougault}
  \affiliation{Normandie Universit\'e, ENSICAEN, UNICAEN, CNRS/IN2P3, LPC Caen, 14000 Caen, France}
  \author{C. Ciampi}
  \affiliation{Dipartimento di Fisica,  Universit\`a di Firenze, Italy} 
  \affiliation{INFN, Sezione di Firenze, Italy} 
\author{J.A. Due\~nas}
\affiliation{Depto. de Ingenier\'ia El\'ectrica y Centro de Estudios Avanzados en F\'isica, Matem\'aticas y
Computaci\'on, Universidad de Huelva, 21007 Huelva, Spain } 
\author{C. Frosin} 
  \affiliation{Dipartimento di Fisica,  Universit\`a di Firenze, Italy} 
  \affiliation{INFN, Sezione di Firenze, Italy} 
  \author{J. D. Frankland}
  \affiliation{GANIL, CEA/DRF-CNRS/IN2P3, 14076 Caen, France}
  \author{D. Gruyer}
  \affiliation{Normandie Universit\'e, ENSICAEN, UNICAEN, CNRS/IN2P3, LPC Caen, 14000 Caen, France}
  \author{N. LeNeindre}
  \affiliation{Normandie Universit\'e, ENSICAEN, UNICAEN, CNRS/IN2P3, LPC Caen, 14000 Caen, France}
  \author{I. Lombardo}
  \affiliation{INFN Sezione di Catania, 95123 Catania, Italy} 
\author{G. Mantovani}
 \affiliation{INFN Laboratori Nazionali di Legnaro, 35020 Legnaro, Italy}
\affiliation{Dipartimento di Fisica, Università di Padova, 35131 Padova, Italy}
\affiliation{Universidade de Santiago de Compostela, 15705 Santiago de Compostela, Spain}
 \author{P. Ottanelli}
  \affiliation{Dipartimento di Fisica,  Universit\`a di Firenze, Italy} 
  \affiliation{INFN, Sezione di Firenze, Italy}   
\author{M. Parlog}
  \affiliation{Normandie Universit\'e, ENSICAEN, UNICAEN, CNRS/IN2P3, LPC Caen, 14000 Caen, France}
  \affiliation{"Horia Hulubei" National Institute of Physics and Nuclear Engineering (IFIN-HH), RO-077125 Bucharest Magurele, Romania}
  \author{G. Pasquali}
\affiliation{Dipartimento di Fisica,  Universit\`a di Firenze, Italy} 
  \affiliation{INFN, Sezione di Firenze, Italy} 
\author{S. Upadhyaya}
\affiliation{Faculty of Physics, Astronomy and Applied Computer Science, Jagiellonian University, 30-348 Kracow, Poland}
\author{S. Valdr\'e}
\affiliation{INFN, Sezione di Firenze, Italy}
\author{G. Verde}
  \affiliation{INFN Sezione di Catania, 95123 Catania, Italy} 
\author{E. Vient}
\affiliation{Normandie Universit\'e, ENSICAEN, UNICAEN, CNRS/IN2P3, LPC Caen, 14000 Caen, France} 
 
\begin{abstract} 

Isospin transport ratio is a powerful method to estimate the neutron-proton (n-p) equilibration  in heavy-ion collisions, and extensively used to obtain information on the asy-stiffness of the nuclear Equation of State. In fact such a ratio is expected to bypass any perturbations introducing a linear transformation of the chosen observable. In particular, it is supposed to overcome contributions
due to emission, either of dynamical or statistical nature, from the primary fragments formed during the collisions. In this paper we explore the validity of this assumption, looking at the quasi-projectile  n-p ratio ($N/Z$) in peripheral and semi-peripheral events for Ca+Ca reactions at 35\amev{}, simulated via the Antisymmetrized Molecular Dynamics transport model, coupled to different statistical decay codes. The statistical de-excitation of the primary fragments introduces a linear transformation at relatively high excitation energies (above 2\amev{}) when the residue approaches the Evaporation Attractor Line, while some
effect is produced at lower excitation
energies due to the occurrence of
some non-linearities. As for fast emissions
after the end of the projectile-target interaction it is shown that they introduce a non-linear transformation too.
\end{abstract}

\maketitle

\section{Introduction}

The isospin transport ratio (also known as imbalance ratio) has been introduced by Rami \textit{et al.}~\cite{bib:rami00_imbalance} in order to extract from experimental data the degree of charge equilibration in heavy-ion collisions in a model independent way. Such a technique exploits combined information from (at least) three systems differing in the neutron-proton ratio $N/Z$: two symmetric reactions, a neutron rich ($NR$) and a neutron deficient ($ND$) one, and an asymmetric system with a neutron content in between that of the two symmetric reactions ($Mix$). Thus, the isospin transport ratio is defined as:
\begin{equation}
R(X)=\frac{2X-X^{NR}-X^{ND}}{X^{NR}-X^{ND}}
\label{eq:ratio}
\end{equation} 
where $X$ is an isospin sensitive observable evaluated in the three systems. 
For the two symmetric systems $R(X)$ is normalized to $+1$ and $-1$ for the n-rich and n-deficient system, respectively. Moreover, 
if the chosen observable linearly depends on the isospin, $R=0$ represents the full n-p equilibration~\cite{bib:rami00_imbalance}.

Such a method has been frequently used in heavy-ion reactions in the Fermi energy domain exploiting different isospin sensitive observables. For instance Ref.~\cite{bib:txliu07_isobaricratio} exploited the $A=7$ mirror nuclei ratio as a function of the rapidity, assessing the transport of isospin asymmetry in Sn+Sn collisions at 50\amev{}. 
This technique is used in the investigation of the asy-stiffness of the nuclear Equation of State (\neos{}), because the charge equilibration degree is influenced by the symmetry energy term~\cite{bib:baran05_eos}. For instance, the isoscaling parameter and  $A=7$ isobaric ratio were used as $X$ and the experimental results were compared with  Boltzmann-Uehling-Uhlenbeck calculations (BUU)~\cite{bib:betty04_isoscaling} or with improved Quantum Molecular Dynamics (iQMD) transport models~\cite{bib:betty09_isoscaling, bib:sun10_isobaricratio}. Although many efforts have been 
done to extract the density dependence of the symmetry energy of the \neos{}, we are still far from a reliable 
determination of the first and higher order coefficients of its Taylor expansion~\cite{bib:magueron18_eos}. In such a scenario it could be useful the direct detection of the isospin content of the Quasi-Projectile (QP) remnant itself~\cite{bib:baran05_transport, bib:napo10_eos}, instead of limiting to its decay products as done in most experiments.  An example in this direction is a recent publication, where May \textit{et al.}  accessed the isospin of the   QP remnant, though they reconstruct it from the produced fragments detected both in charge and mass by means of the NIMROD multi-detector~\cite{bib:may19_ratio}.

The use of the isospin transport ratio to estimate the isospin equilibration presents many advantages. According to Rami \textit{et al}~\cite{bib:rami00_imbalance}, if the three reactions are investigated under identical experimental conditions, the ratio is insensitive  to systematic uncertainties due to the apparatus; the errors are essentially statistical. Isospin transport ratio is also expected to largely remove the sensitivity to fast dynamical emissions, secondary decays and Coulomb effects~\cite{bib:betty04_isoscaling, bib:txliu07_isobaricratio, bib:betty09_isoscaling, bib:coupland11_imbalance_effects}. 
More generally, this method bypasses any effect which introduces a linear transformation $F_{L}$ on the adopted observable $X$, \ie{} $R(F_{L}(X)) = R(X)$. For this to be possible, the transformation $F_{L}$ must be applied to all the reactions but it can depend on the ordering variable, used to follow the evolution of $R$ with the impact parameter or on the phase space subset under investigation. Instead, non-linear transformations $F_{NL}$ introduce a deformation of the isospin transport ratio, therefore  $R(F_{NL}(X))\neq  R(X)$. 

In this paper we aim at investigating, in binary dissipative collisions, the effects that particle emissions (fast or evaporative) from primary fragments  introduce on the isospin transport ratio itself through model calculations.
The influence of quantities more related to the
nuclear interaction such as mean field, in-medium cross section and dynamical cluster production has been studied in Ref.~\cite{bib:coupland11_imbalance_effects}, where an investigation 
via BUU model is reported; indeed, to our knowledge, a clear information on the effects of the statistical de-excitation of the fragments and fast emission is still lacking. 

In the framework of models used to simulate nuclear collisions, it
is quite common and convenient~\cite{bib:defilippo05_dynamicalfission, bib:defilippo12_timescale, bib:piantelli19_fiasco, bib:csym16, bib:tian17_cc, bib:tian18_cc} to assume a two step process: 
a dynamical phase, described by a transport model, and the following statistical de-excitation of the produced hot fragments performed by means of a statistical decay code. In this work we chose to adopt the Antisymmetrized Molecular Dynamics (AMD) model~\cite{bib:amd92} to describe the dynamical evolution,  since it has been shown to be able to predict in a reliable way the main features of the collisions in the Fermi energy domain~\cite{bib:ono19_rev}, also in semi-peripheral collisions~\cite{bib:camaiani18_iwm, bib:piantelli19_fiasco, bib:piantelli19_isofazia, bib:tian17_cc, bib:tian18_cc}. Three different decay models have been used as afterburner, \ie{} two different versions of GEMINI statistical code (GEMINI++~\cite{bib:gemini} and GEMINIf90~\cite{bib:geminif90}), and SIMON~\cite{bib:simon}, the afterburner associated with the HIPSE event generator~\cite{bib:lacroix04_hipse, bib:lacroix05_hipse}. In particular, differences between GEMINIf90 and GEMINI++ were recently observed when used as afterburner of the same dynamical code~\cite{bib:piantelli19_fiasco, bib:piantelli19_isofazia}; on the other hand, differently from GEMINI, SIMON takes into account account Coulomb trajectories during the decays~\cite{bib:simon}. Concerning the investigation of fast emission (predicted within AMD), we adopt a time back-tracing procedure, presented in Ref.~\cite{bib:piantelli19_isofazia}, in order to characterize the events 
as a function of the collision  time; this allows us to directly access the produced hot Quasi-Projectile and Quasi-Target (QT) nuclei just after the end of the interaction phase.

This paper is organized as follows. In sec.~\ref{sec:app} the experimental context and the simulation codes will be presented. Sec.~\ref{sec:evap} is dedicated to the investigation of the effects of the statistical de-excitation on the isospin transport ratio, while sec.~\ref{sec:pre} will focus on fast emission contributions. Finally in sec.~\ref{sec:concl} conclusions are drawn.

\section{The experimental context and the simulation codes}
\label{sec:app}

As test bench we chose a set of reactions involving Ca ions, which present a rather wide range of stable isotopes and thus  are usable in conventional experiments. 
In fact the present study was motivated by recent experimental
data obtained by our groups in particular those of  the INDRA+VAMOS campaign~\cite{bib:phd_quentin, bib:wigg12_caca, bib:boisjoli12_caca} and one of the first FAZIA experiment dedicated 
to the investigation of isospin effects in reactions with Ca ions~\cite{bib:camaiani18_iwm, bib:phd_camaiani}. 
Consequently in this paper we considered the following systems: 
the symmetric  n-rich and n-deficient $^{48,40}$Ca+$^{48,40}$Ca reactions, used as references for the isospin transport ratio, and the asymmetric one $^{48}$Ca+$^{40}$Ca, where isospin diffusion acts.
All the calculations have been done at 35\amev{}. We chose to select the $N/Z$ of the QP as isospin sensitive observable, since it is  expected to be a good tool to investigate the asy-stiffness of the \neos{}~\cite{bib:baran05_transport, bib:napo10_eos} and the $N/Z$ of the QP remnant can be also measured by means of detectors as FAZIA~\cite{bib:bougault14, bib:valdre18} and VAMOS~\cite{bib:vamos, bib:vamos++} characterized by high isotopic separation capability.
For this reason we will focus on binary reactions, \ie{} peripheral and semi-peripheral events ($b_{red}=b/b_{gr}\geq 0.4$, where $b_{gr}$ is the grazing impact parameter). 

As already anticipated for the simulation of the dynamical phase of such reactions we adopted the AMD model. A complete description of the AMD transport code can be found elsewhere~\cite{bib:ono19_rev, bib:amd92, bib:amd96, bib:amd99, bib:amd_proc, bib:amd16_pion}. Here, we remind that AMD is  based on molecular dynamics where a system of nucleons is described by a Slater determinant of Gaussian wave packets. The time evolution 
of the system is obtained by means of a time dependent variational principle, taking into account both mean field contribution and two-nucleon collision processes. The mean field is described via the effective interaction Skyrme SLy4~\cite{bib:skyrme}, using 
$K_{sat}=230\,$MeV for the incompressibility modulus of the nuclear matter and $\rho_0=0.16\,$fm$^{-3}$ for the saturation density. Two different parametrizations for the symmetry energy can be selected. A soft symmetry energy corresponds to a symmetry energy term (zero order term) of $E_{sym}=32\,$MeV and to a first order parameter $L=46\,$MeV; a stiff symmetry energy ($L=108\,$MeV) can be obtained by changing the density dependent term in the SLy4
force~\cite{bib:amd16_pion}. Such recipes 
are compatible with the reported values for realistic
parametrizations~\cite{bib:magueron18_eos}. In this work we focus on the asy-stiff parametrization, except where otherwise stated.  
Two-nucleon collisions are implemented as stochastic transitions within AMD states under the constraint of momentum
and energy conservation and the strict fulfillment of the
Pauli principle. The transition probability depends on the in-medium nucleon-nucleon cross section, which can be considered, within some limits, as a free parameter of the model. In this work we adopted the parametrization proposed in Ref.~\cite{bib:lopez14_inmedium}, with a screening factor of $y=0.85$. In order to take into account cluster correlations arising during the dynamics, cluster states are included among the possible achievable final states~\cite{bib:amd_proc, bib:piantelli19_fiasco}. It is important to note that the present investigation is not focused on the fine tuning of the parameters within the AMD model. We remind that the aim of this work is to understand how fast emissions and statistical evaporation act on the isospin transport ratio, thus affecting (or not) the estimation of the n-p equilibration degree.

For each system we produced approximately 40000 events with a triangular distribution of the impact parameter up to the grazing value, stopping the AMD code at 500$\,$fm/c from the onset of the interaction: this is a sufficiently long time to assure that the dynamical phase is concluded, when the primary ejectiles have reached the thermodynamical equilibrium, and to ensure that the fragment mutual Coulomb repulsion is negligible~\cite{bib:piantelli19_fiasco}.

The hot QP nuclei produced at 500$\,$fm/c have been used as inputs to different statistical decay codes. For each primary event, 100 secondary events have been produced for GEMINI++ (GEM++), GEMINIf90 (GEMf90) and SIMON, in order to estimate the effects on the n-p equilibration produced by different statistical codes. 

The event selection is performed at the end of the statistical stage, as for the experimental data; the QP and QT remnants from binary collisions are selected as $Z\geq$12, only accompanied  by neutrons, Light Charged Particles (H and He ions) and Intermediate Mass Fragments, produced during the decay path. The selected sample represents 62\% of the whole statistics, and 98\% of the events in the selected range of centrality ($b_{red} \geq 0.4$).

\section{Statistical de-excitation effects}
\label{sec:evap}

\begin{figure}
   \centering 
   \includegraphics[width=1\columnwidth]{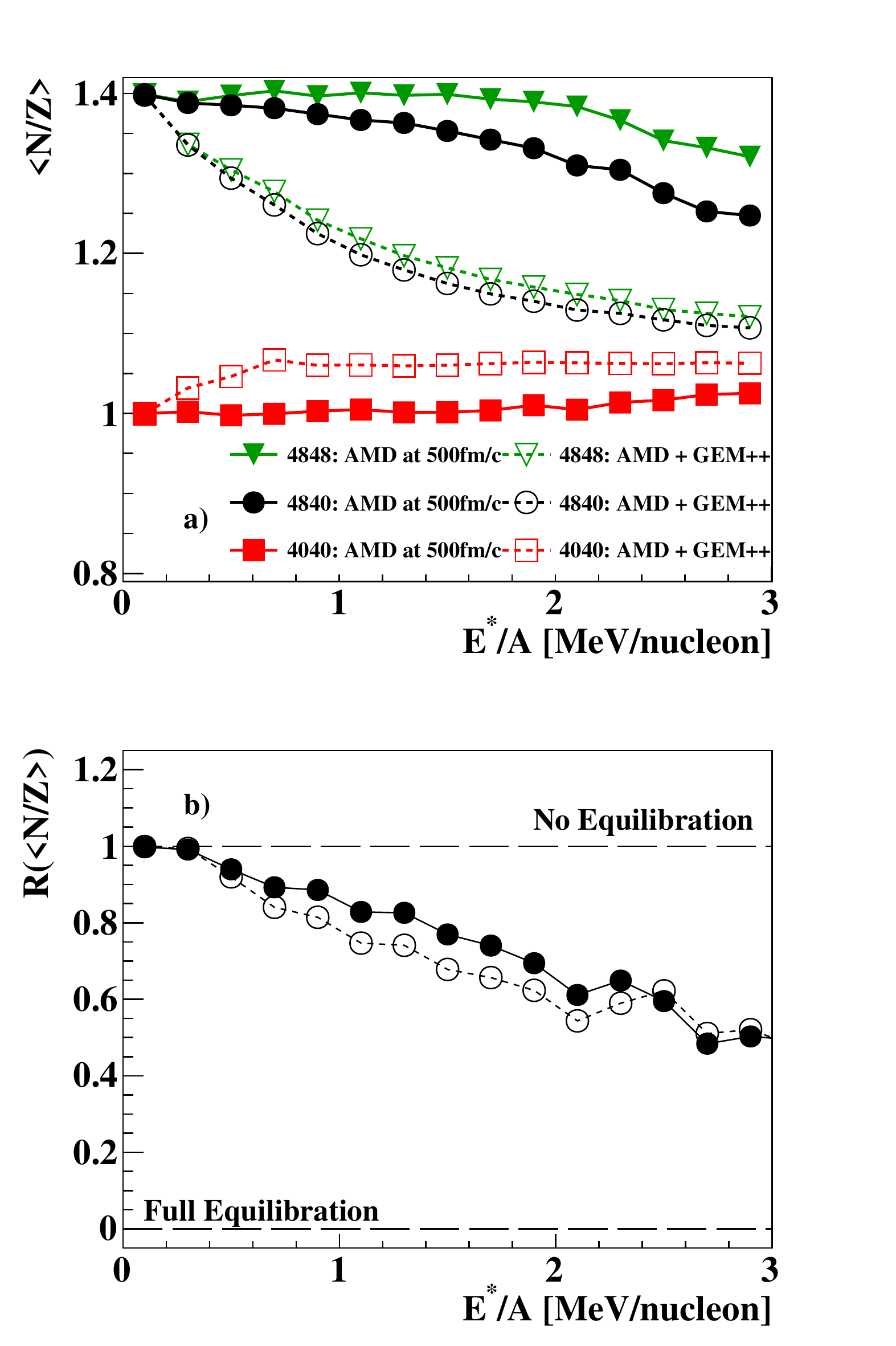} 
   \caption{(Color Online) a) Evolution of the neutron-to-proton ratio of the QP as a function of the QP excitation energy per nucleon calculated at 500$\,$fm/c. Solid symbols refer to the QP at 500$\,$fm/c, open symbols to the  QP remnant, \ie{} at the end of the statistical de-excitation. b) Isospin transport ratio, of the \caa{}+\ca{}, QP as a function of its excitation energy per nucleon measured at 500$\,$fm/c. Solid circles refer to primary QP at 500$\,$fm/c, open circles correspond to the QP remnant after the statistical de-excitation. Lines are drawn to guide the eyes.}  
   \label{fig:nsuz} 
\end{figure}

\begin{figure*}
   \centering 
   \includegraphics[width=1\textwidth]{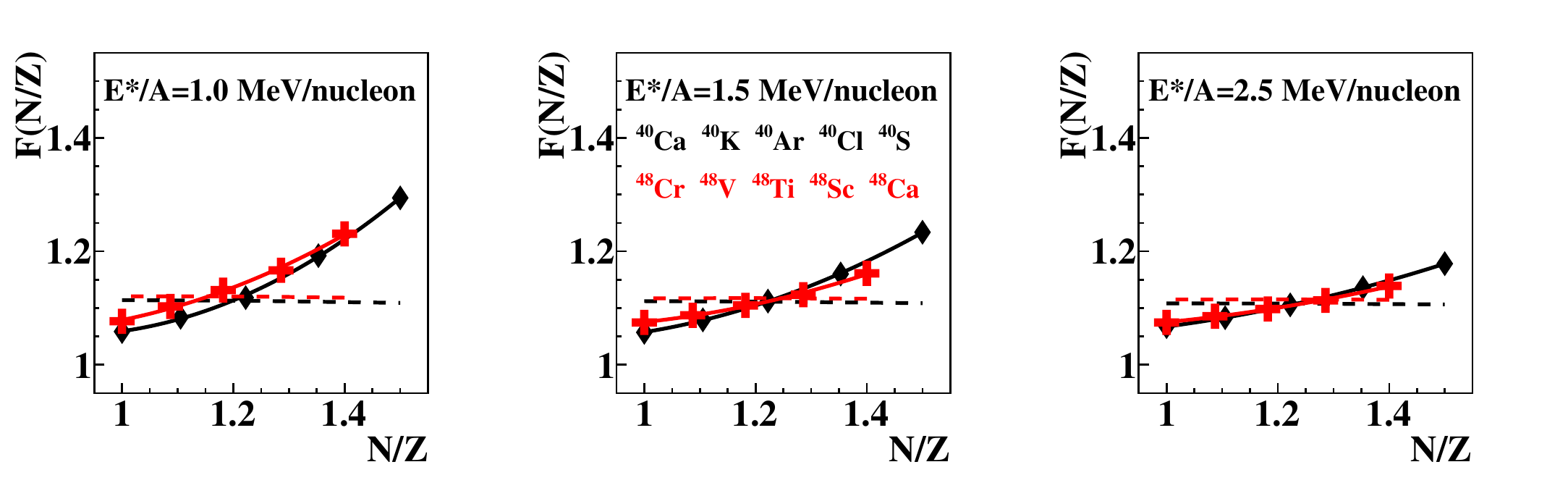} 
   \caption{(Color Online) GEMINI++ simulation for two different sets of nuclei, $A=40$ (solid black diamond) and $A=48$ (solid red crosses). In the picture correlations between the neutron-to-proton ratio obtained at the end of the evaporation ($F(N/Z)$) and that of the input nuclei ($N/Z$) are shown, for three different values of \estara{} as quoted. Black and red dashed lines represent the value predeicted by the EAL\cite{bib:eal}, for the $A=40$ and $A=48$ nuclei, respectively. }  
   \label{fig:test} 
\end{figure*} 

In this Section we explore the effects of the statistical emission on charge equilibration evaluated via isospin transport ratio. In this sense, in order to distinguish the effect of the statistical decay from that of a more prompt emission, we can  compare the equilibration degree obtained at the end of the afterburner with that from primary QP at 500$\,$fm/c (the end of the dynamical phase according to our modelization). In this way any difference is attributable to the  statistical decay only. However, we stress that the "true" charge equilibration degree is pertaining to the system at the end of the projectile-target interaction and any subsequent emission may perturb it, whatever its nature. 

Fig.~\ref{fig:nsuz}(a) shows the evolution of the average
n-p ratio \nsuz{} of the QP as a function of its excitation energy per nucleon ($E^{*}/A$) evaluated at 500$\,$fm/c. 
For each bin of $E^{*}/A$, containing $M$ fragments, the \nsuz{} is calculated as follows:
\begin{equation}
\langle \frac{N}{Z} \rangle = \frac{\sum_j \sum_i \frac{N_i(j)}{Z_j} Y_i(j)}{\sum_j \sum_i Y_i(j)}
\label{eq:nsuz}
\end{equation}
where  $N_i(j)$ and $Y_i(j)$ represent the neutron number and the yield of the $i$-th isotope of the $j$-th element with charge number $Z_j$; in particular $M=\sum_j \sum_i Y_i(j)$.
We used \estara{} as order variable since it is one of the main parameters which govern the statistical decay of a nucleus; 
we underline that \estara{} scales as a function of the impact parameter; moving from lower to higher values, the events are ordered from peripheral to more central events. The \caa{}+\caa{} reaction is represented by green triangles, \caa{}+\ca{} by black circles and \ca{}+\ca{} by red squares. Solid symbols refer to the primary QP at 500$\,$fm/c and open symbols correspond to the QP remnant. 
As the excitation energy increases (\ie{} from peripheral to more central collisions) the \nsuz{} of the primary QP moves from the projectile value (1.4 and 1 for \caa{} and \ca{}, respectively), decreasing in the n-rich and mixed 
reactions, slightly increasing  in the n-deficient one. The systems present a clear hierarchy. The effect of the isospin diffusion can be seen as the differences of the \nsuz{} values in the \caa{}+\ca{} system with respect to the symmetric n-rich one~\cite{bib:defilippo12_timescale}. The afterburner strongly modifies the values of \nsuz{} but the system hierarchy survives. 

\begin{figure}
   \centering 
   \includegraphics[width=1\columnwidth]{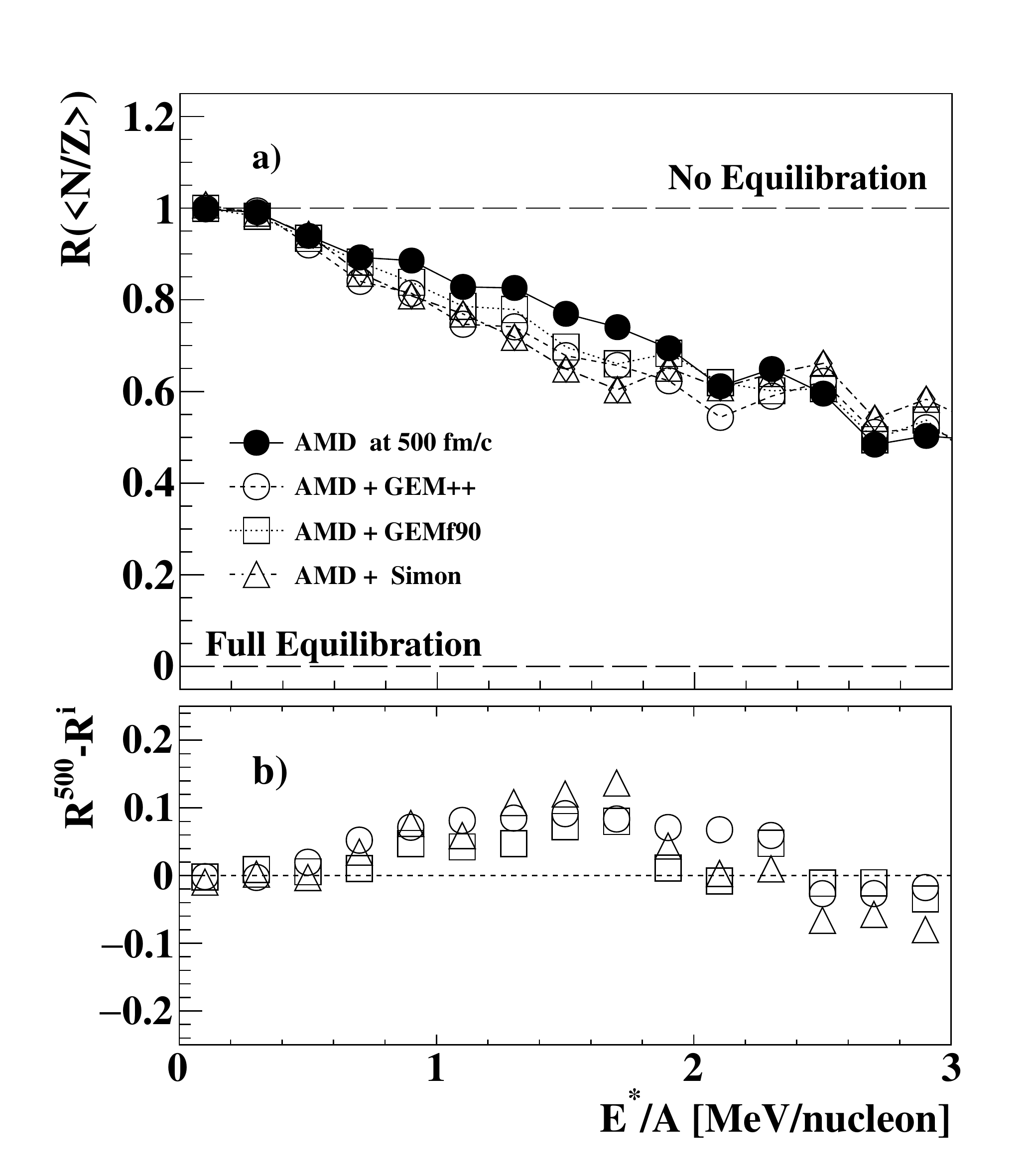} 
   \caption{Panel a) Isospin transport ratio for the AMD primary QP at 500$\,$fm/c (solid circles) and at the end of the statistical de-excitation. Panel b) Differences in the equilibration degree of the QP remnant with respect to that obtained at 500$\,$fm/c (open symbols). Statistical errors are smaller than the marker size. Lines are drawn to guide the eyes.}  
   \label{fig:all} 
\end{figure} 

Fig.~\ref{fig:nsuz}(b) shows the corresponding isospin transport ratio obtained using as $X$ the values of \nsuz{} presented in Fig.~\ref{fig:nsuz}(a). Both at 500$\,$fm/c and at the end of the statistical decay, the system evolves from $R=1$ (no equilibration) 
downwards as the \estara{} increases. As expected, the system evolves towards the charge equilibrium once the collision becomes more dissipative~\cite{bib:baran05_eos, bib:baran05_transport, bib:coupland11_imbalance_effects}. However, some differences between the values at 500$\,$fm/c and those at the end of the de-excitation appear.
We observe a discrepancy of about 0.1 in the range 0.6$\div$2.2$\,$MeV/nucleon. Such finding may suggest a non-linear effect introduced by GEMINI++, otherwise its contribution should be removed by the isospin transport ratio.

By means of GEMINI++ we tested this hypothesis. 
Let's define as $F$ the transformation of $N/Z$ representing the effects of the evaporation. $F$ is a function of the neutron number $N$, atomic number $Z$, angular momentum $J$, level density parameter $a$, and excitation energy per nucleon \estara{}, \ie{} the inputs that rule the statistical decay of a nucleus~\cite{bib:geminif90, bib:gemini}. We produced two sets of GEMINI++ simulations for several specific hot nuclei fixing their mass: the first with $A=40$ nuclei, the second with $A=48$ nuclei. The chosen nuclei are labeled in the central panel of Fig.~\ref{fig:test}(b). The spin of such nuclei is fixed at 6$\,\hbar$, according to the average value predicted by AMD at 500$\,$fm/c. In order to test the nature of the transformation introduced by GEMINI++ we show the $N/Z$ of the input nuclei versus that obtained (on average) from the evaporation residue at the end of the decay ($F(N/Z)$). Solid black diamonds are for the $A=40$ nuclei, solid red crosses for the $A=48$ nuclei.  For each nucleus we computed 10000 events.

Results are reported in Fig.~\ref{fig:test}(a,b,c), for three values of \estara{} as quoted in the panels. 
It is likely true that the effect of the secondary decay is non-linear below 2.5\amev{}. The curvature of these relations is consistent with the differences of the  isospin transport ratio in Fig.~\ref{fig:nsuz}(b): a positive curvature produces a decrease of $R$, while a negative one an increase (see Sec.~\ref{sec:pre}). 
However, the transformation tends to be linear with increasing excitation energy. Such behavior can be interpreted taking into account the Evaporation Attractor Line (EAL)~\cite{bib:eal},\ie{} the locus in the $Z-N$ plane which is approached by the nuclear residues after the evaporation decay and which runs close to the $\beta-$stability line. In particular, the higher the excitation energy of the parent nucleus, the closer the final value residue to the EAL.
The $N/Z$ values predicted for the EAL for the tested input nuclei are quite similar and are represented with a dashed red (black) line for $A=48$ ($40$). As expected, $F(N/Z)$ approaches the EAL values as \estara{} increases. Consequently, using the EAL as a reference locus, the following conclusions can be drawn. The transformation $F$ introduced by the evaporation is linear once the evaporation residues are close to the EAL; differently, for low values of \estara{}, $F$ has a non-linear behavior. Such non-linearity is reflected in the isospin transport ratio of Fig.~\ref{fig:nsuz}(b), causing a discrepancy between the ratios calculated at 500$\,$fm/c or for post-evaporative fragments. It follows that (at least for this region of nuclei close to Ca) for  \estara{}$\leq 2\,$MeV/nucleon, the obtained isospin transport ratio is affected by the statistical decay.

The investigation of the causes behind the non-linearity at low \estara{} is out of the goal of this paper. A possible contribution could be due to structure effects, since is well known that they affect the particle emission at low excitation energy~\cite{bib:bruno19, bib:more19, bib:camaiani18}. In this sense, the isoscaling analysis~\cite{bib:betty04_isoscaling, bib:betty09_isoscaling} could be more suited as it divides out many structure effects to the first order.

We tested the robustness of the previous results changing the afterburner either using a different version of GEMINI~\cite{bib:geminif90}, or using a completely independent code (SIMON)~\cite{bib:simon}. Results are reported in Fig.~\ref{fig:all}(a); also results of Fig.~\ref{fig:nsuz}(b) are shown for sake of comparison. The difference with respect to the degree of equilibration  measured at 500$\,$fm/c ($R^{500} - R^{i}$, where $i$ represents the various afterburners) is presented in Fig.~\ref{fig:all}(b).
The obtained trends for GEMINIf90 and SIMON statistical codes
are comparable with that obtained for GEMINI++. Each afterburner introduces similar distortions of the isospin transport ratio, in a similar range of \estara{}, with respect to the charge equilibration evaluated at the start of the afterburner application. However, a closer look shows that GEMINIf90 gives results closer to the \mbox{n-p} equilibration at 500$\,$fm/c. Above 2\amev{}, the linearity seems to be recovered, and the differences fluctuate around zero.

Therefore, the conclusion of this section can be summarized as follows. At high excitation energies, the decaying nuclei approach the EAL and the transformation introduced by the evaporation on the $N/Z$ is practically linear; instead, for lower excitation energies, non-linearities develop and cause distortions of the isospin transport ratio. This is confirmed for the three used decay models although the effects produced by GEMINIf90 are weaker. In future,  such a study can be extended to other isospin sensitive observables used as $X$ (eq.\ref{eq:ratio}), as the $\alpha$ isoscaling parameter.

\section{Fast emission effects}
\label{sec:pre}

\begin{figure}
   \centering 
   \includegraphics[width=1\columnwidth]{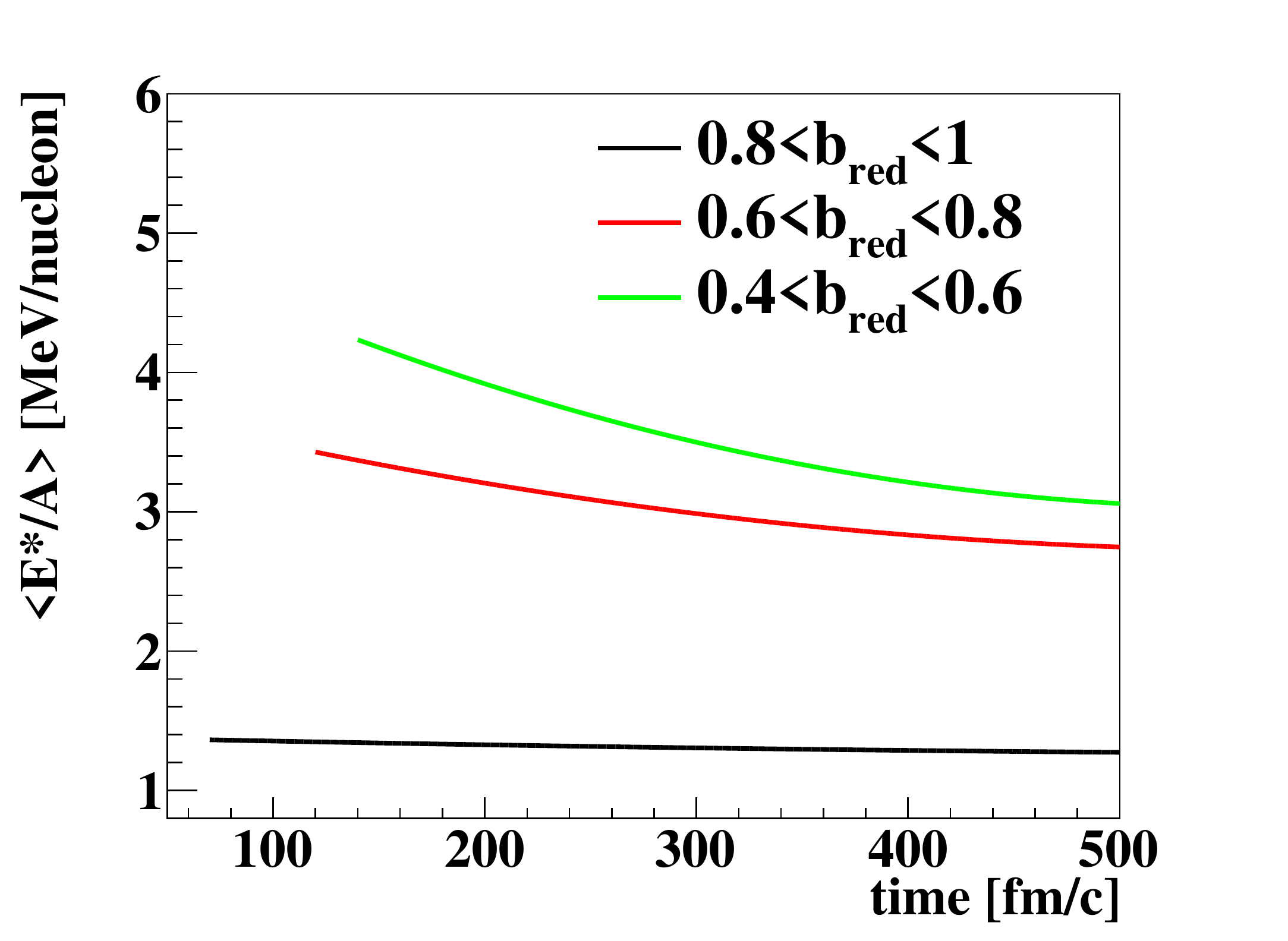} 
   \caption{(Color Online) AMD predictions for the \caa{}+\ca{} system: excitation energy per nucleon as a function of time, for three bins of centrality according to the legend. Each line starts at the average QP-QT separation time in the selected range of impact parameters.}  
   \label{fig:ex} 
\end{figure} 

\begin{figure}
   \centering 
   \includegraphics[width=1\columnwidth]{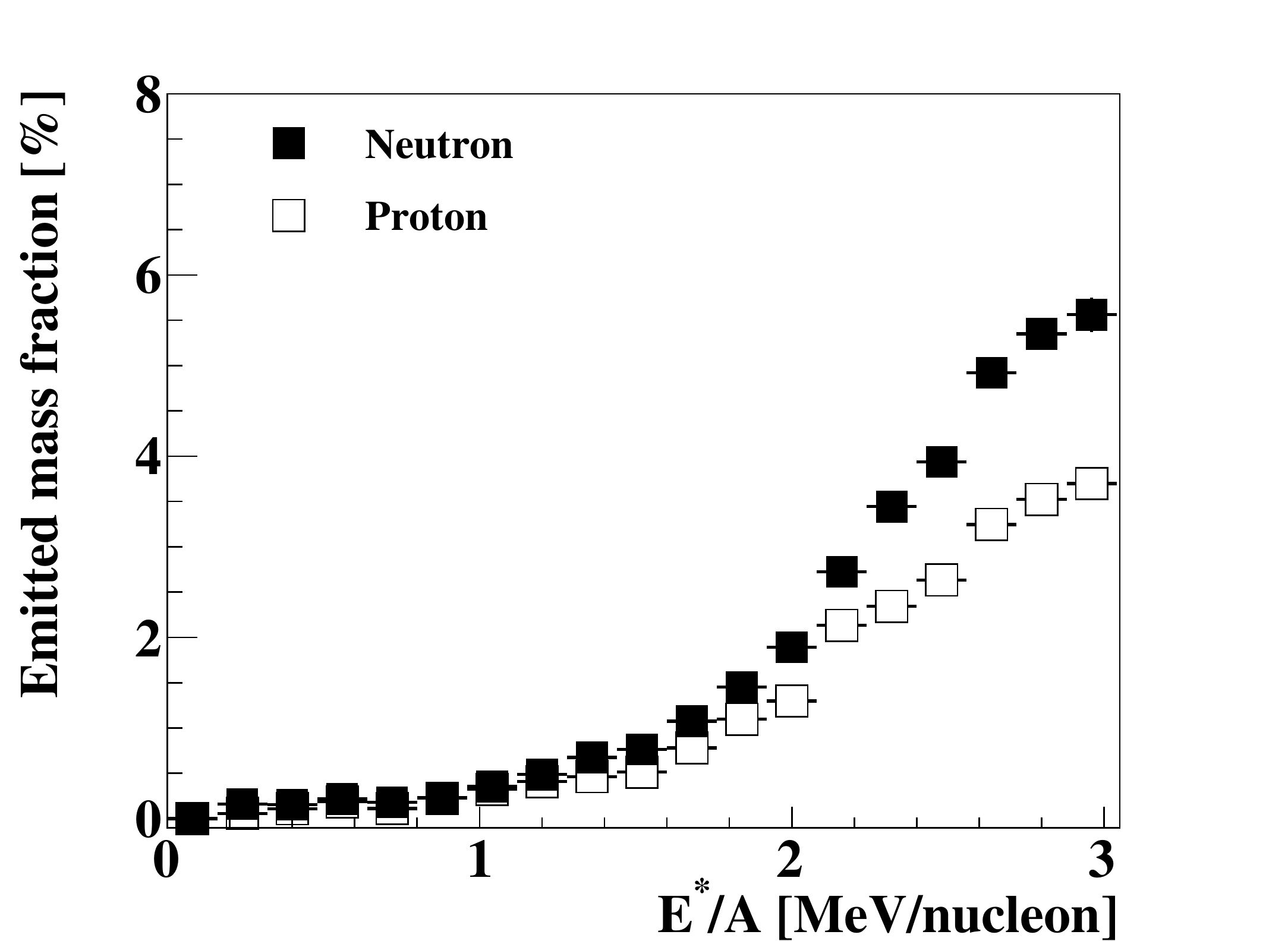} 
   \caption{Percentage of neutrons (solid square) and protons (open square) emitted in the whole phase-space between $t_{DIC}$ and 500$\,$fm/c in the \caa{}+\ca{} system. The neutron (proton) number of the system has been chosen as normalization factor.}
   \label{fig:pre_em} 
\end{figure}

\begin{figure}
   \centering 
   \includegraphics[width=1\columnwidth]{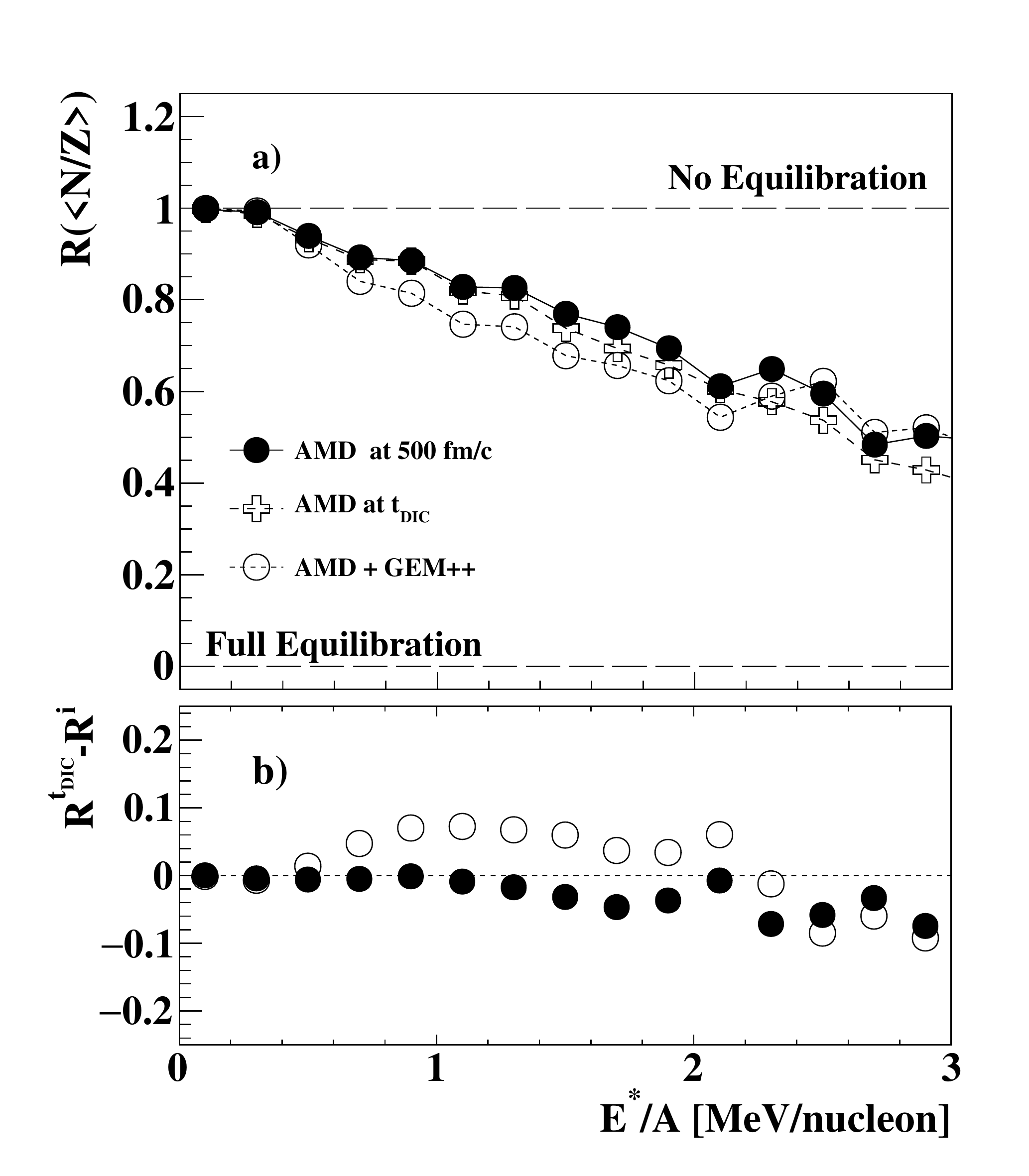} 
   \caption{Panel a) Isospin transport ratio for the AMD primary QP at 500$\,$fm/c (solid circles), at the end of the statistical de-excitation (empty circles), and at the end of the interaction $t_{DIC}$ (empty crosses).  Lines are drawn to guide the eyes. Panel b) Differences in the equilibration degree of the QP at 500$\,$fm/c (solid circle) and at the end of the statistical de-excitation (empty circle) with respect to that obtained at the end of the interaction. Statistical errors are smaller than the marker size. 
   }  
   \label{fig:tsplit} 
\end{figure}

\begin{figure*}
   \centering 
   \includegraphics[width=1\textwidth]{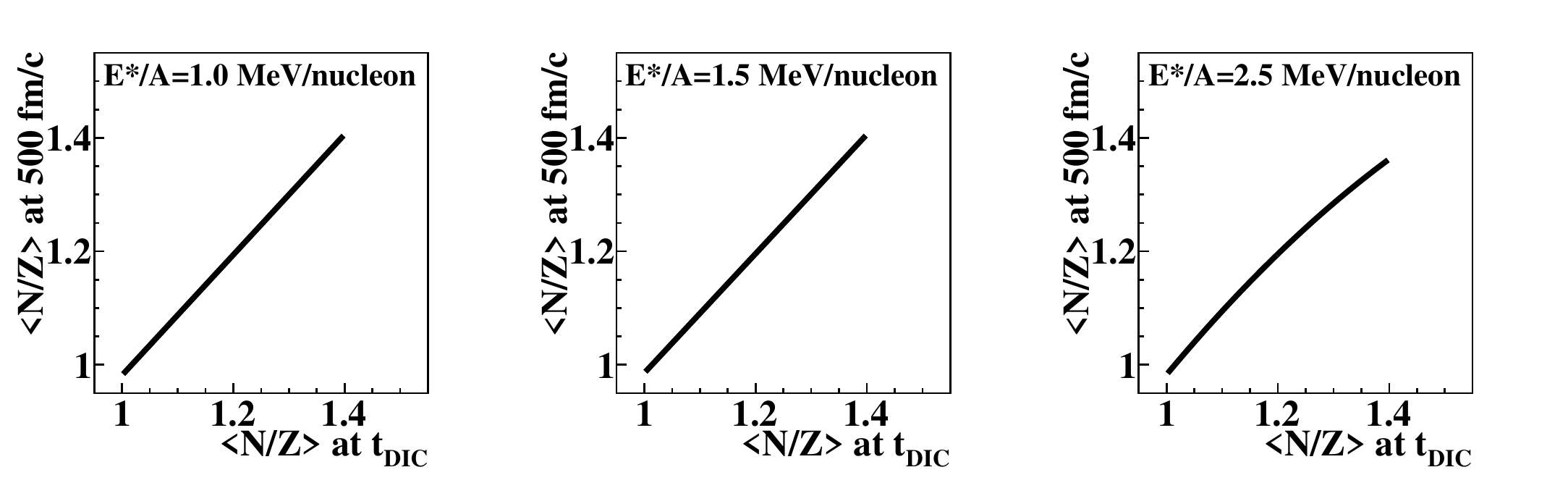} 
   \caption{Average transformations on \nsuz{} due to the emission between $t_{DIC}$ and 500$\,$fm/c for three bins of $E^{*}/A$. The selection on \estara{} has been done at 500$\,$fm/c.}  
   \label{fig:testpre} 
\end{figure*}

We move now to investigate the influence of the emissions before 500$\,$fm/c, \ie{} those predicted by the dynamical code, on the isospin transport ratio. In particular we focus on the emissions that occur between the QP and QT reseparation time ($t_{DIC}$) and 500$\,$fm/c. As anticipated in Sec.~\ref{sec:evap}, we want to stress that n-p exchanges between projectile and target stops at $t_{DIC}$ and any subsequent emission may produce a distortion on the correct estimation of the equilibration degree reached by the system at $t_{DIC}$. In order to access to QP at $t_{DIC}$ we adopted the same procedure described in Ref.~\cite{bib:piantelli19_isofazia}. 
For each event selected at the end of the statistical stage, we applied the AMD fragment recognition algorithm every 20$\,$fm/c  from 500$\,$fm/c to the onset of the interaction; two wave packets (nucleons) are taken as belonging to the same fragment if the distance between their centers is within 5$\,$fm. We go back in time until a unique system with a mass and charge comparable with the interacting projectile-target system is found, thus defining the splitting time $t_{DIC}$. For better accuracy and consistency, going back in time we require 
that for each step the size of the identified QP be not less than that at the previous time step. In particular, 
since in this paper we are dealing with binary events, the fragments at $t_{DIC}$ correspond to the QP residues at the end of the statistical stage.

Fig.~\ref{fig:ex} shows the evolution of the average excitation energy per nucleon ($\langle E^{*}/A \rangle$) of QP as a function of time for three different bins of centrality for the \caa{}+\ca{} system.
For the investigated system, AMD predicts a split time of approximately $50-150\,$fm/c, of course depending on the impact parameter.  Each line starts at the average QP-QT separation time according to the selected range of centrality

Once the excited QP emerges after the collision it undergoes a series of particle emissions which continues up to 500$\,$fm/c. The nature of the particle emissions modeled by AMD after the interaction is not clearly assessed. Just after $t_{DIC}$, the QP is  not yet equilibrated, e.g. having largely deformed shape and large-amplitude collective motions. Consequently, some emissions  can be ascribable to non-equilibrium effects. On the other hand, as the time increase, the primary fragments tend to the thermodynamical  equilibrium and part of the observed emissons may have a more statistical nature. Indeed, the properties of such emissions  can be consistent with a statistical description~\cite{bib:futura09, bib:futura06, bib:ono96_eq, bib:ono96_eq2}, even if they are  calculated within a dynamical model and not in the typical Hauser-Feshbach scheme.

Fig.~\ref{fig:ex} also shows that the emissions after the 
reseparation increase with the violence of the collisions. 
This indeed is demonstrated by the decrease of \estara{} starting from the maximum value around $t_{DIC}$. 
We see that the excitation energy negligibly decreases for peripheral 
collisions while it varies of about 1$\,$MeV/nucleon for the most central considered events.

In this sense, Fig.~\ref{fig:pre_em} shows the evolution as a function of \estara{} of the number of neutrons (protons) emitted between $t_{DIC}$, and 500 fm/c, normalized to the number of neutrons (protons) of the whole \caa{}+\ca{} system  with solid (open) symbols. Below 2\amev{} the percentage of fast emitted nucleons is relatively small, while it becomes significant at higher excitation energy.

The comparison between the degree of equilibration (evaluated from the isospin transport ratio) at 500$\,$fm/c and $t_{DIC}$ is finally reported in Fig.~\ref{fig:tsplit}(a), with solid black circles and open stars, respectively; for sake of comparison also the charge equilibrium at the end of the AMD+GEMINI++ calculation is reported. Fig.~\ref{fig:tsplit}(b) shows the differences with respect to the the equilibration degree at the end of the interaction phase ($R^{t_{DIC}}$). For sake of homogeneity 
with the previous section we chose again \estara{} as ordering variable, keeping in mind that it also reflects the centrality of the collision, from peripheral to more central events as \estara{} increases. The trend as a function of \estara{} is very similar between the equilibration degree at $t_{DIC}$ and at 500$\,$fm/c, and some slight differences arise for $E^{*}/A>2\,$MeV/nucleon (see Fig.~\ref{fig:tsplit}(a,b)).

Again, this result suggests a non-linear transformation 
introduced by the fast emissions between $t_{DIC}$ and 500$\,$fm/c. 
With respect to the previous section, here it is not possible to select a single source at $t_{DIC}$ and correlate it with the corresponding value at 500$\,$fm/c. 
However, the average trend of the transformation 
can be calculated correlating the \nsuz{} of the QP at 500$\,$fm/c ($F(\langle N/Z \rangle)$) with the original value at $t_{DIC}$: each panel of Fig.~\ref{fig:testpre} shows the obtained trend for three bins of \estara{}. As the excitation energy per nucleon increases, fast emissions introduce  non-linear distortions that cannot be fully recovered by the isospin transport ratio. Indeed, for peripheral collisions corresponding to lower excitation energy, the "fast emission" contributions is negligible (see. Fig.~\ref{fig:pre_em}).

\section{Summary and conclusions}
\label{sec:concl}

In this paper we investigated the effects on the isospin transport ratio of all particle emissions occurring after the end of the interaction within a model scheme.

We performed our calculation in Ca+Ca systems at 35\amev{}, thus scanning a large range of the neutron-proton ratio, from 1.0 for the \ca{}+\ca{} system, up to 1.4 in the \caa{}+\caa{}. The choice of 
the reactions is strictly related to the experiment performed by the FAZIA collaboration aiming at a detailed study of the isospin equilibration in the \caa{}+\ca{} system by means of the isospin transport ratio. We chose to adopt a standard two step approach: a first dynamical phase followed by a statistical de-excitation of the primary fragments. The AMD transport model~\cite{bib:ono19_rev} describes the dynamical evolution of the collision up to 500$\,$fm/c, and then different statistical codes have been used as afterburners: GEMINI++~\cite{bib:gemini}, GEMINIf90~\cite{bib:geminif90} and SIMON~\cite{bib:simon}. As isospin sensitive observable used to calculate the isospin transport ratio we 
exploited the neutron-proton ratio of the quasi-projectile, since it is expected to be a good probe to test the asy-stiffness of the nuclear Equation of State~\cite{bib:baran05_transport, bib:napo10_eos}. As a consequence, we focused on peripheral and semi-peripheral reactions ($b_{red}>0.4$).

The main findings of this work are the following.
The statistical de-excitation produces a transformation of 
the QP $N/Z$ values which tends to be linear at relatively high excitation energies when the residues approach the Evaporator Attractor Line~\cite{bib:eal}. 
In the present case this corresponds to excitation energies above 2$\,$MeV/nucleon. The non-linearity developed at lower \estara{} causes a perturbation in the $N/Z$ that cannot be fully canceled by the 
isospin transport ratio. These variations are almost the same within the three tested statistical codes, being weaker using the GEMINIf90 version.

The particle emission just after the QP-QT separation are still 
described by the AMD code itself. They occur all along before the 
(arbitrary fixed) end of the dynamical phase (500$\,$fm/c) when the pure statistical 
code is switched on as afterburner. In this time interval the nucleon exchange process between projectile and target is exhausted but such  emissions may distort the isospin equilibration signal. These emissions produced  by the AMD code are not purely statistical, \ie{} not calculated following the Hauser-Feshbach  approach. As a matter of fact they perturb the real n-p equilibration degree at an extent that increases with the collision violence (decrease of the impact parameter). In particular, their 
effect on the isospin transport ratio becomes significant for 
semi-central events, approximately for $\langle b_{red} \rangle < 0.8$. For more peripheral reactions, at least for these reactions, the fast emissions are negligible (and thus their effect on the isospin transport ratio).

\begin{figure}
   \centering 
   \includegraphics[width=1\columnwidth]{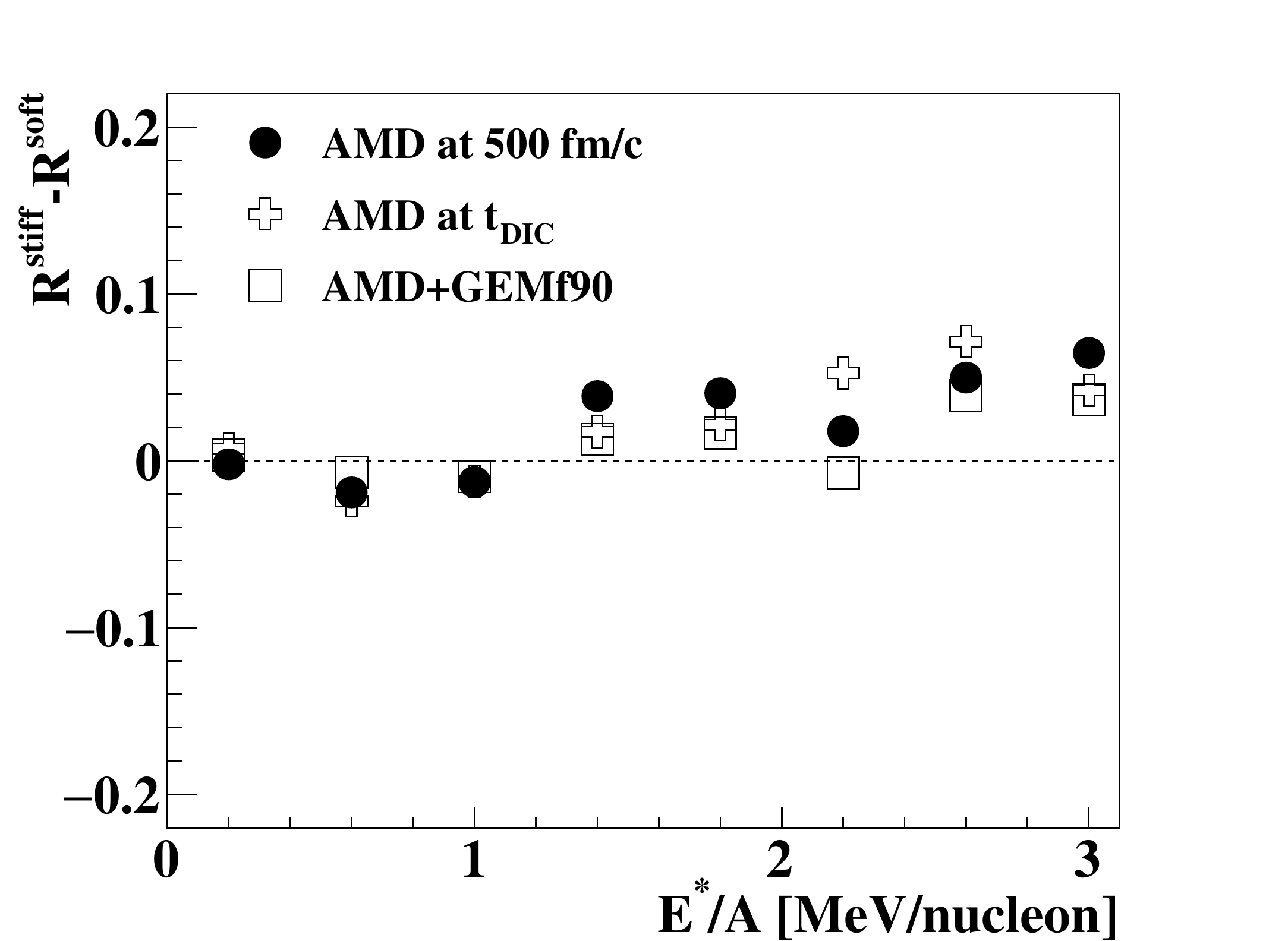} 
   \caption{AMD calculations for \caa{}+\ca{} system at 35\amev{}. Differences in the n-p equilibration between the asy-stiff and asy-soft recipes via isospin transport ratio. Results at 500$\,$fm/c and at the end of the statistical decay performed via GEMINIf90 are shown. Bin number has been reduced, with respect to previous figures, to prevent fluctuations.}  
   \label{fig:stiffsoft} 
\end{figure} 
In conclusion, the investigation presented in this paper has shown that both the statistical de-excitation of primary fragments and the "fast dynamical" emissions can influence the observed n-p equilibration via isospin transport ratio.  
The overall effects can be seen comparing the equilibration degree at the separation time with that at the end of the statistical stage (see fig.~\ref{fig:tsplit}(a,b)). They introduce a non-linear distortion on the \nsuz{} of the QP in two distinctive range of centrality, approximately above and below $\langle b_{red} \rangle \approx 0.8$ for the statistical and dynamical emission, respectively. Both effects have to be taken into account once a comparison with experimental data, aiming at constraining the asy-stiffness of the \neos{}, is performed. In fact, the expected differences in the isospin transport ratio  between asy-stiff and asy-soft recipes might be small due to the presence of couteracting effects~\cite{bib:coupland11_imbalance_effects} as for instance the development of cluster correlations during the dynamical phase~\cite{bib:amd16_pion}. Consequently, they could be of the same order of the distortion introduced by secondary decays and fast dynamical emissions. For the considered Ca-system, this is shown in Fig.~\ref{fig:stiffsoft}, where the differences of the n-p equilibration  between the asy-stiff ($R^{stiff}$) and asy-soft ($R^{soft}$) recipe are shown, both at the end of the interaction ($t_{DIC}$), at 500$\,$fm/c and at the end of the statistical stage via GEMINIf90.

\begin{acknowledgements}

This work required the use of a lot of computation
time for the production of the simulated data. We would 
like to thank the GARR Consortium for the kind use
of the cloud computing infrastructure on the platform
cloud.garr.it. We would like to thank also the INFN-CNAF for the use of its cloud computing infrastructure. Moreover, A. Ono was supported by JSPS KAKENHI Grant No. JP17K05432. Finally, we would like to warmly thank Prof. Giacomo Poggi for very helpful suggestions and discussions.

\end{acknowledgements}

\bibliography{biblio.bib} 

\end{document}